\documentclass[runningheads]{llncs}
\usepackage{graphicx}
\newcommand{\Lagr}{\mathcal{L}}
\usepackage{hyperref}

\makeatletter
\newcommand{\printfnsymbol}[1]{%
  \textsuperscript{\@fnsymbol{#1}}%
}
\makeatother
\begin{document}
\title{Segmentation-free Estimation of Aortic Diameters from MRI Using Deep Learning
}
\titlerunning{ }
%
\author{
Axel Aguerreberry \inst{1}
\and Ezequiel de la Rosa \inst{2}
\and Alain Lalande \inst{3, 4}\thanks{These authors contributed equally.} \and Elmer Fernández \inst{1, 5}\printfnsymbol{1} 
}
\authorrunning{ }
%
\institute{School of Exact, Physical and Natural Sciences, National University of Cordoba, Cordoba, Argentina.
\email{axel.aguerreberry@gmail.com}\\
\and Department of Computer Science, Technical University of Munich, Munich, Germany.
\and ImViA laboratory, University of Burgundy, Dijon, France.
\and Medical Imaging Department, University Hospital of Dijon, Dijon, France.
\and CIDIE - CONICET, Catholic University of Cordoba, Cordoba, Argentina.
}
\maketitle          
\begin{abstract}
Accurate and reproducible measurements of the aortic diameters are crucial for the diagnosis of cardiovascular diseases and for therapeutic decision making. Currently, these measurements are manually performed by healthcare professionals, being time consuming, highly variable, and suffering from lack of reproducibility. In
this work we propose a supervised deep-learning method for the direct estimation of aortic diameters. The approach is devised and tested over 100 magnetic resonance angiography scans without contrast agent. All data was expert-annotated at six aortic locations typically used in clinical practice.
Our approach makes use of a 3D+2D convolutional neural network (CNN) that takes as input a 3D scan and outputs the aortic diameter at a given location. In a 5-fold cross-validation comparison against a fully 3D CNN and against a 3D multiresolution CNN, our approach
was consistently superior in predicting the aortic diameters. Overall, the 3D+2D CNN achieved a mean absolute error between 2.2-2.4 $mm$ depending on the considered aortic location. These errors are less than 1 $mm$ higher than the inter-observer variability. Thus, suggesting that our method makes predictions almost reaching the expert's performance. We conclude that the work allows to further explore automatic algorithms for direct estimation of anatomical structures without the necessity of a segmentation step. It also opens possibilities for the
automation of cardiovascular measurements in clinical settings.


\keywords{Aortic Diameter  \and Magnetic Resonance Imaging \and Convolutional
Neural Networks.} \end{abstract}

\section{Introduction}
The determination of the maximum aortic diameters using imaging techniques is crucial for cardiovascular diseases (such as aortic dissection, aortic hematoma, aortic aneurysm and coarctation of the aorta) assessment. The diagnosis and monitoring of these diseases require precise and reproducible measurements of the  aortic diameters for optimal treatment decision making \cite{goldstein2015multimodality}. Maximum aortic diameters are crucial not only for deciding surgical procedures, but are also crucial at postoperative stages, where an increase in the aortic diameter can eventually lead to complications.

Currently, aortic markers are manually delineated by physicians, being time demanding and suffering from observers variability and from poor
reproducibility. The inter-observer measurement variability of proximal aortic diameters range from 1.6 to 5 $mm$ \cite{goldstein2015multimodality}. Annotations are usually done using 3D imaging software, generally integrated in the scanner's workstations. Given the complexity of the aortic anatomy, a consensus has been established for defining where these annotations should be conducted by defining a set of reference points. 

Traditional approaches for automatically estimating the aortic diameters require, as the first step, the segmentation of the vessel, such as the work of Suzuki et al. \cite{suzuki2018automated}. Segmentation is typically performed following supervised learning approaches \cite{duquette20123d}, which require full expert delineation of the aorta. The process is not only time demanding, but also observer-biased. As a consequence, the segmentation performance affects the diameters' measurement. Different segmentation-free approaches have been proposed to estimate cardiac parameters. For instance, Xu et al. directly estimate the cardiac cavity, the myocardial area, and the regional wall thickness in cine-MRI sequences \cite{xue2017direct,xue2018full}. Debus and Ferrante make use of a 3D spatio-temporal CNN for regressing left ventricle volumes from MRI scans \cite{debus2018left}. More recently, Zhang et al. proposed a model for the direct estimation of coronary artery diameter in X-ray angiograhpy images \cite{zhang2019direct}. It consists of a multi-view module with attention mechanisms for learning spatio-temporal coronary features. Furthermore, machine learning approaches have also been explored for the direct estimation of cardiovascular markers. For instance, Zhen et al. conducts direct and joint bi-ventricular volume estimation by extracting multiple image features and fitting a random forest regression algorithm \cite{zhen2014direct}.

In this work, an alternative strategy is proposed for directly estimating
aortic diameters bypassing segmentation. We propose a 3D+2D CNN that inputs MRI angiography scans and estimates the maximum aortic diameter at different locations.

\section{Materials and methods}

\subsection{Dataset}
The dataset consists of 100 magnetic resonance angiography studies without contrast agent. A free-breathing, ECG-triggered steady-state free precession sequence was employed. The studies have been acquired with a 1.5 T scanner (Siemens Healthineers, Erlangen, Germany) at Dijon University Hospital (France). The voxel resolution varies among scans in between 1.22x1.22x1.50 $mm^{3}$ and 1.56x1.56x1.50 $mm^{3}$.

We follow the conventional clinical process to evaluate and diagnose the aorta.
The database annotations have been performed by several different physicians of the hospital, where each scan has been annotated a single time by one of the experts. For each scan, six aortic diameters typically used in clinical practice \cite{mongeon2016multimodality} have been measured: the diameter at the i) Valsalva sinuses, ii) sinotubular junction, iii) ascending aorta, iv) horizontal aorta, v) aortic isthmus and vi) thoracic (descending) aorta. The annotations were done using the 3D data by means of the software $syngo.via$ (Siemens Healthineers, Erlangen, Germany). We directly retrieved the MRI scans with their corresponding diameter measurements.
It is worth to mention that only the scalar of these measurements was available. Thus, the ground-truth does not include coordinate points or location information of the measurements.
Our database included pathological cases with varying severity, with not only normal cases (89\%, $n$=89), but also cases with aortic dissection (4\%, $n$=4) and presence of prosthesis (7\%, $n$=7).
Moreover, with the aim of assessing the inter-observer variability for each aortic diameter, a randomly chosen 20\% of the database has been re-annotated by an expert in the field (AL), which was blinded to the first observer results.

\subsection{Proposed approach}
We propose an end-to-end framework for automatically regressing the aortic diameters from MR angiography data. The approach receives as input the 3D angiography scans and outputs the measured aortic diameter at a specific location. It is worth pointing out that no aortic segmentation mask is required. The same framework and training strategy is used for each of the different aortic diameters.

\subsection{Preprocessing}
All images undergo a reslicing and affine registration step for homogenizing the scans resolution and for aligning and/or orienting all scans respectively. Images are resliced to a voxel resolution of a 1x1x1 $mm^{3}$. This step, which is done automatically, is crucial for allowing the neural network to properly learn the voxel resolution scale and, hence, properly predict the aortic diameters. The registration is performed by aligning each scan to a single reference image, accounting the reference one with i) high voxel resolution and ii) good image quality (qualitatively assessed). The usage of a single reference scan is enough for our purposes, since we only aim to compensate for acquisition orientation differences across scans.

After image registration, with the aim of reducing GPU memory usage, we automatically select the 16 central slices of the scan which include the whole aorta anatomy. After that, the images are normalized following a z-score normalization.

\subsection{CNN Architecture}
\begin{figure}[h]
			
			\includegraphics[width=\textwidth]{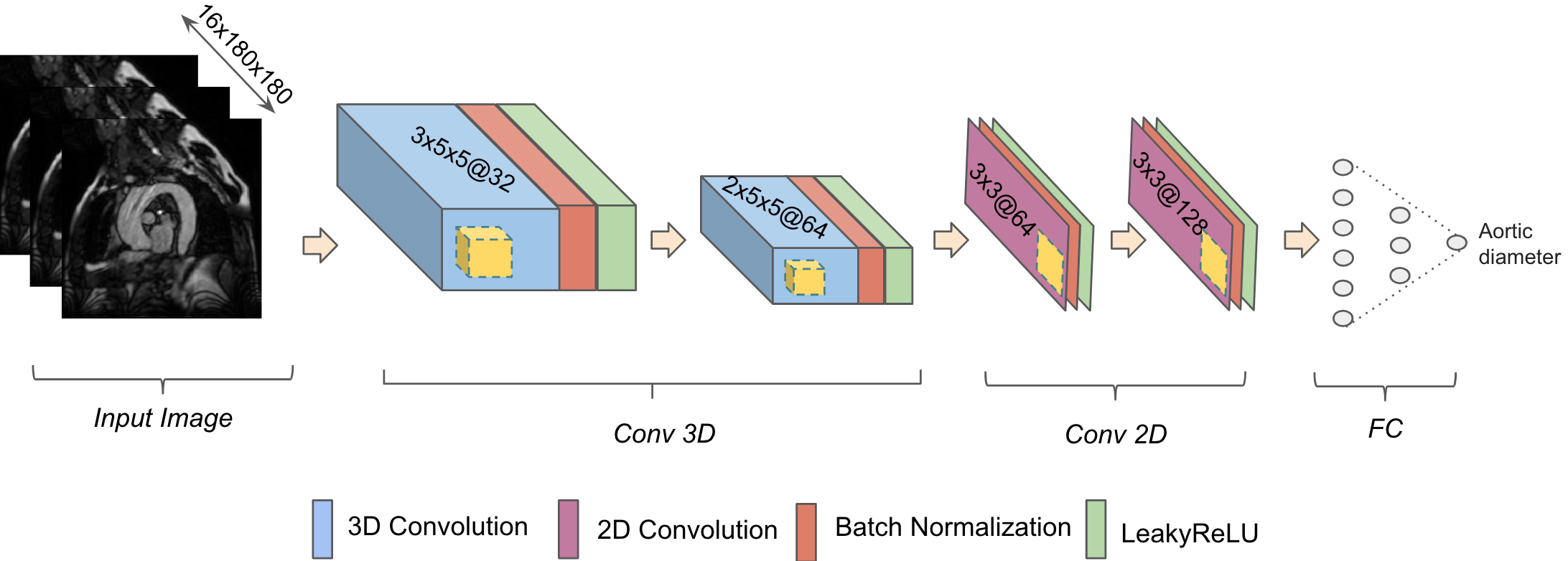}	
			\caption{\scriptsize Proposed CNN architecture. It consists on a 3D + 2D network. Inside each convolution block are displayed the amount and size of the kernels. Conv: convolutional layer. FC: Fully-connected layer.}
			\label{fig:2d3d}
\end{figure}

An overview of the proposed CNN architecture is presented in Figure \ref{fig:2d3d}.
The network takes sequences of 16 MRI slices as input and outputs the corresponding aortic diameter at a given landmark. The architecture is fairly straightforward: it consists on 3D convolutional operations, followed by 2D convolutional operations with fully connected layers at the end. The last neuron represents the aortic diameter estimated (there is one specific network for each measurement). $LeakyReLU$ activation functions were used across all layers, except for the last fully-connected layer, where linear functions are preferred, in order to map the neural network output to the expected diameter values. Besides, for faster convergence of the model, we include batch-normalization layers after each convolutional operation.

\subsection{Training procedure}
We train the proposed network by minimizing a mean-squared-error loss function. Given an MRI scan $x$ and a ground-truth diameter $y$ (for the sake of simplicity, we consider $y$ to represent any of the six aortic diameters), the loss function is defined as:

\smallskip
\begin{equation}
\Lagr(y, \hat{y}) = \frac{1}{N}\sum_{i=1}^{N}(y_i - \hat{y_i})^{2}
\label{eq:funcion_perdida}
\end{equation}

where $y_i$ and $\hat{y_i}$ represent the predicted and ground-truth aortic diameters respectively. Optimization of Equation \ref{eq:funcion_perdida} was conducted using Adam optimizer with mini-batches of size equal to eight.

For regularizing the model we make use of traditional data augmentation transformations (such as Gaussian noise addition and slight image rotations to compensate for inaccuracies during registration). Besides, we introduce a L2 regularization term to Equation \ref{eq:funcion_perdida}. The L2 weighting factor (known as $\lambda$ in literature) was empirically set to 0.1. Dropout layers \cite{srivastava2014dropout} were only included after the fully connected layers (dropout factor $p=0.5$).

\subsection{Comparison against other methods}
To the best of our knowledge this is the first approach addressing direct aortic diameters estimation. As such, comparing our proposal against reference approaches is not straightforward. We propose two baseline models for comparing our method with: a 3D regression CNN and a 3D multi-resolution CNN. 

\subsubsection{3D Regression CNN.}
This network is a fully 3D version of our 3D+2D proposal. All convolutional operations are 3D. We try to keep the network as close as possible to our proposal. Thus, the amount of parameters was kept similar to the 3D+2D CNN. Activation functions are $LeakyRelu$ for all layers except for the last fully connected layer.

\subsubsection{3D Multiresolution CNN.}
We also implemented and trained the multiresolution CNN  proposed by Tousignant et al.\cite{tousignant2019prediction}. The approach inputs 3D MRI data and outputs a scalar. Though the model was originally devised for predicting disability outcome in patients with multiple sclerosis, its architecture can be used for addressing our problem. Our adapted implementation consists of three consecutive 3D convolutional blocks, followed by a fully connected layer. Each convolutional block contains parallel convolutional pathways. Thus, the input to the convolutional block is split into four pathways with different kernel sizes. Finally, the pathways are concatenated and fed as the new input to the consecutive block. 

\subsection{Evaluation criteria}
In order to evaluate the performance of the proposed method, we conduct 5-fold cross-validation experiments for each aortic diameter. As performance metrics we use the mean squared error (Equation \ref{eq:funcion_perdida}), the mean absolute error ($MAE=\frac{1}{N}\sum_{i=1}^N\mid y_{i} - \widehat{y_{i}}\mid$), Pearson correlation coefficient (PCC) and Bland-Altman analysis \cite{bland1986statistical}.





\section{Results and discussion}
All experiments have been performed in a 12 GB GPU with processor  Intel Core i7-7700. Training of the model took on average $\sim$ 3.9 hours. While predicting one aortic diameter took $\sim$ 8.34 $\mu$s, performing its manual annotation took $\sim$ 50 s per diameter per scan.  

In this work we assume that the same model and training strategy could be shared across the different aortic diameters. So, for selecting the optimal model and for conducting hyper-parameter tuning, we first experiment taking into account the annotation at the level of the Valsalva sinuses. This diameter was chosen given the fact that is generally a challenging measurement in clinical practice and, typically, suffers from higher inter-observer variability according to the experts. Once the optimal model and training strategy have been defined, the analysis was extended by training and validating the model over all the remaining aortic landmarks (i.e. sinotubular junction, ascending aorta, etc). 

\subsection{Effect of the registration approach}

We have compared three different preprocessing pipelines using: i) No registration, ii) $rigid$ registration and iii) $affine$ registration. The results are summarized in Table \ref{tablaregistration}, where it can be appreciated the advantage of using affine registrations in the pipeline.

\begin{table}[h]
                 \centering
                \caption{Prediction performance for different registration approaches trained at the Valsalva sinuses' diameter using the model 3D+2D. MAE: Mean absolute error; MSE: mean squared error; PCC: Pearson correlation coefficient.}
                \smallskip
                \label{tablaregistration}
                \begin{tabular}{cccr}
                    \hline
                    Registration & MAE ($mm$) &MSE ($mm^{2}$)& PCC \\ \hline     	
                    \textbf{Affine} & \textbf{2.341}  &  \textbf{8.721} &  \textbf{0.914} \\ 
                    Rigid  & 2.839  & 13.951 & 0.886 \\
                    No registration & 4.549 & 32.825 & 0.778 \\
                    \hline
                \end{tabular}
\end{table}

\subsection{Effect of the chosen architecture}
In Table \ref{tabla1} the performance obtained for each architecture trained at the level of the Valsalva sinuses are shown.

\begin{table}[h]
                 \centering
                \caption{Prediction performance for the three compared architectures  trained at the Valsalva sinuses' diameter. MAE: Mean absolute error; MSE: mean squared error; PCC: Pearson correlation coefficient.}
                \smallskip
                \label{tabla1}
                \begin{tabular}{cccr}
                    \hline
                    Model & MAE ($mm$) &MSE ($mm^{2}$)& PCC \\ \hline  	
                    3D & 3.493  & 18.151 & 0.859 \\
                    \textbf{3D+2D} & \textbf{2.341}  &  \textbf{8.721} &  \textbf{0.914} \\
                    Multiresolution & 4.269 & 30.342 & 0.785 \\
                    \hline
                \end{tabular}
\end{table}

The CNN made up of 3D+2D convolutional layers was consistently better than the other models, by reaching the lowest MAE, the lowest MSE and the highest PCC. 
These results show that incorporating 2D convolutions after the 3D model improves the overall performance.

\subsection{Performance at each aortic location}
\begin{figure}[hbt!]
			
			\includegraphics[width=\textwidth]{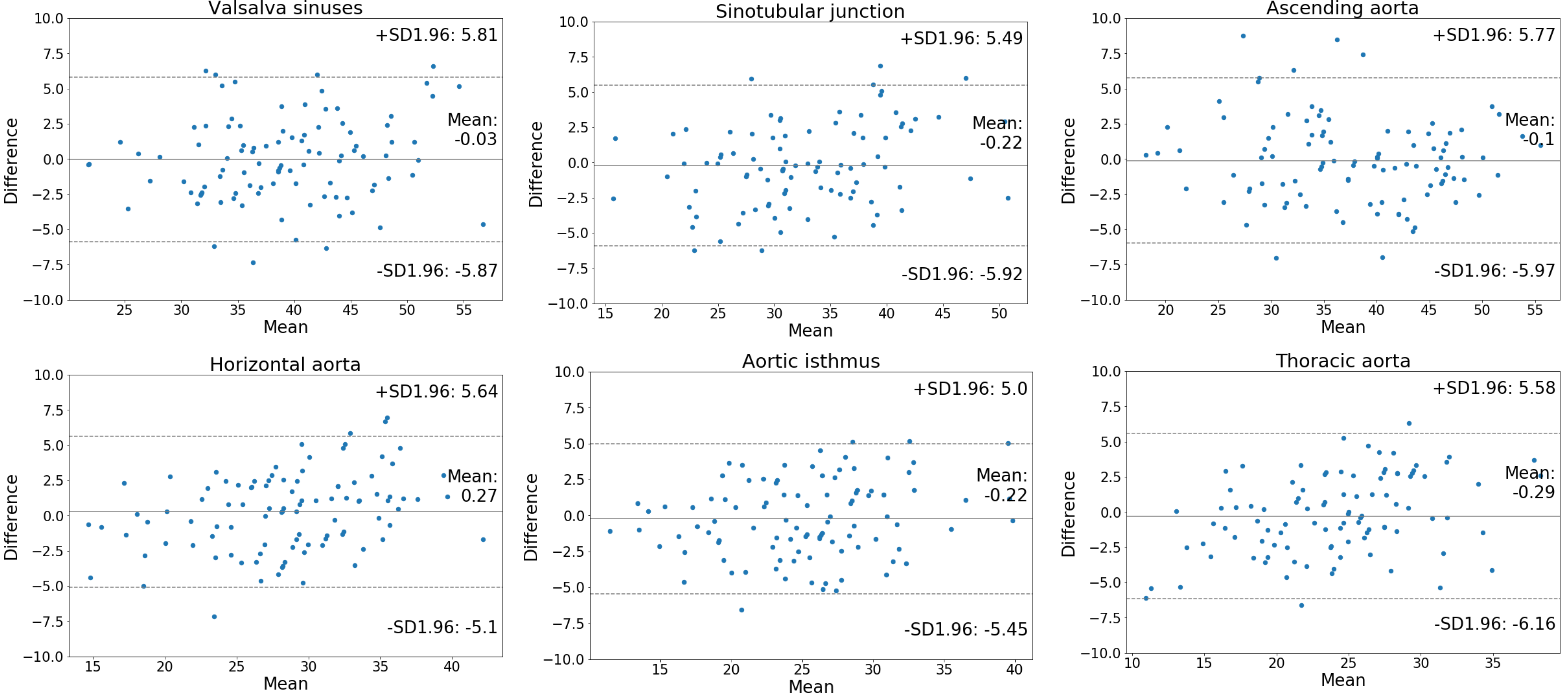}	
			\caption{\scriptsize Bland-Altman plots for each given landmark.}
			\label{fig:BA}
\end{figure}

Table \ref{tabla2} summarizes the inter-observer errors and the performance obtained at each aortic location when using the model and training strategy defined above. The prediction errors for each landmark are very similar, with tiny differences that can be attributed to the complexity of the aortic anatomy at the considered location. Overall, our model predicted with a MAE in between 2.2-2.4 $mm$. This is less than 1 mm of MAE difference when compared against the experts, which obtained a MAE in between 1.3-1.7 $mm$. The lowest prediction and inter-observer errors correspond to the diameters at the aortic isthmus location. The highest MAE at prediction time were obtained for the thoracic descending aorta. The Bland Altman plots (Fig. \ref{fig:BA}) show a low mean bias at all locations without a clear trend towards under/over-estimation of the diameters.

\begin{table}[h]
                \centering
                \caption{Results at each aortic location.}
                \smallskip
                \label{tabla2}
                \begin{tabular}{ccccccc}
                    \hline
                    & \multicolumn{3}{c}{Predicted} & \multicolumn{3}{c}{Inter-observer}\\ 
                    \hline
                    Landmark & MAE ($mm$) &MSE ($mm^{2}$)& PCC & MAE ($mm$) &MSE ($mm^{2}$) & PCC\\ \hline     	
                    Valsalva sinuses & 2.341 & 8.721 & 0.914 & 1.65 & 3.95 &0.961\\ 
                    Sinotubular junction & 2.316 & 8.652 & 0.923 & 1.5 & 3.26 & 0.980\\
                    Ascending aorta & 2.283 & 8.245 & 0.927 & 1.75 & 4.26&0.989\\ 
                    Horizontal aorta & 2.270 & 7.832 & 0.913 & 1.55 & 7.05& 0.913\\
                    Aortic isthmus & 2.251 & 7.364 & 0.912 & 1.35 & 2.52&0.956\\
                    Thoracic aorta & 2.435 & 9.162 & 0.893 & 1.45 & 3.00&0.940\\
                    \hline
                \end{tabular}
\end{table}

\section{Conclusions}
In this work we propose an end-to-end, segmentation-free 3D+2D CNN for the direct estimation of aortic diameters. To our knowledge, this is the first work attempting to regress aortic biomarkers bypassing the anatomical aorta segmentation. The method outperforms a 3D CNN and a multiresolution 3D CNN in terms of MAE, MSE and PCC. The inclusion of a 2D CNN in the model showed some sort of information gain when compared against fully 3D CNNs. The proposed model achieved performances  almost at the inter-observer variability error. For all the considered aortic diameters, the CNN obtained a MAE less than 1 mm compared to the experts variability (i.e., errors on average up to one voxel of difference). 

This work allows to continue researching intelligent algorithms devised to directly estimate anatomical structures without image delineation. The method is general and could be tested over different image modalities,  different anatomical structures and markers. Particularly, for the problem addressed in this work, there is much more effort to consider towards its translation to clinical settings. We aim to improve the model performance by making use of problem-specific data augmentation. Random scaling at image and ground truth levels could help to simulate anatomical size variations. Moreover, devising deep-learning interpretable models is a big milestone for this project. We aim to allow anatomical localization into the CNN, such that clinicians could get some knowledge regarding where the measurements have been performed. 
Another approach could be to compare our method with a landmark detection model. 
For instance, Vlozntzos et al. \cite{vlontzos2019multiple} proposed the use of multi-agent reinforcement learning for landmark localization. They proposed a detection approach for multiple landmarks based on a Deep Q-Network architecture.
Currently, since we do not have landmark annotations in our database, but only the measurements retrieved from the clinical reports, these approaches are not applicable to our problem.
However, landmark localization through reinforcement learning are valid approaches for future research. Finally, another future perspective is exploring multi-task learning models for performing multiple aortic location estimations with a single model.

\bibliographystyle{splncs04}

\end{document}